%Paper: cond-mat/9210019
%From: lopez%uimrl.dnet@uimrl7.mrl.uiuc.edu
%Date: Wed, 21 Oct 92 11:56:48 -0500

\input tex.tex

\rightline{October 9, 1992}

\title  Response functions and Spectrum of Collective Excitations of
Fractional Quantum Hall Effect Systems.
\author Ana Lopez \qquad and \qquad Eduardo Fradkin
\affil Department of Physics
       University of Illinois at Urbana-Champaign
       1110 W.Green St., Urbana IL 61801

\abstract{ We calculate the electromagnetic response functions of a
Fractional Quantum Hall system within the framework of the fermion Chern-Simons
theory for the Fractional Hall Effect (FQHE) which we developed before.
Our results are valid in a semiclassical expansion around the average field
approximation (AFA). We reexamine the AFA and the role of
fluctuations. We argue that, order-\-by-\-order in
the semiclassical expansion, the response functions obey the correct
symmetry properties required by Galilean and Gauge Invariance and by
the incompressibility of the fluid. In particular, we find that the
low-momentum limit of the semiclassical approximation to the response
functions is exact and that it saturates the $f$-sum rule.
We obtain the spectrum of collective excitations of FQHE systems
in the low-momentum limit. We find a rich spectrum of modes which
includes a host of quasiparticle-quasihole bound states and, in general,
two collective modes coalescing at the cyclotron frequency. The Hall
conductance is obtained from the current-density correlation
function, and it has the correct value already at the semiclassical level.
We applied these results to the problem of the screening of
external charges and fluxes by the electron fluid, and obtained
asymptotic expressions of the charge and current density profiles, for
different types of interactions.
Finally, we reconsider the anyon superfluid within our scheme and derive
the spectrum of collective modes for interacting hard-core bosons and
semions. In addition to the gapless phase mode, we find a set of gapped
collective modes.}

\body
\endtopmatter

\head{I.Introduction}

The physical systems which exhibit the Fractional Quantum Hall Effect
(FQHE) present a very rich response to external electromagnetic
perturbations. While some of the observed phenomena, such as cyclotron
resonance, can be understood in terms of simple global motions of the
center of mass under the combined influence of electric and magnetic
fields, the spectrum of collective excitations is certainly determined
by the interactions. Given the unusual features of the Laughlin states
and its generalizations, it is expected that the same features should
largely determine the behavior of the collective modes too. However, in
spite of the great progress that has been made in the understanding of
the ground state, a general theory of the electromagnetic response
functions and of the spectrum of collective modes,
valid for all the incompressible states, has been lacking. This is the
main motivation and goal of this paper.

Various theoretical approaches have been proposed to explain the
FQHE.
The Laughlin-Haldane-Halperin \refto{laughlin1,hald,halp} approach is based
on the Laughlin variational ansatz for the ground state wave function.
The Laughlin wave function gives the correct value for the Hall conductance,
and yields an excellent ground state energy \refto{laughlin1,hald}.
Later on, Halperin \refto{halp} realized that the quasiparticles supported by
this state exhibit not only fractional charge but that they are {\it anyons},
particles with fractional statistics\refto{wilczek1}.
A hierarchy of {\it daughter} states at other fractions different from the
fundamental fractions, \ie $\nu = {1\over m}$, can be constructed by
considering a Laughlin-type ground state of the fractionally charged
quasiparticles defined relative to the parent state one step up in the
hierarchy. The higher-order FQHE states occur at a sequence of rational
filling fractions.

Related to this approach is the composite fermion theory of the FQHE developed
 by Jain \refto{jain1}. He found that the low energy states of the FQHE
can be described in terms
of weakly interacting composite fermions, where a composite fermion is an
electron bound to an even number of vortices. He also proposed simple
Jastrow-Slater
trial wavefunctions for the incompressible FQHE states as well as for their low
energy excitations. The validity of these wavefunctions was confirmed by
calculating numerically their overlap with the true Coulomb states for
systems with small number of particles.

Another approach consists on an effective Landau-Ginzburg field theory for
the  \break
FQHE \refto{zhang, zhang1}. It was shown that the mean field solutions and the
small fluctuations of the Landau-Ginzburg effective action give a correct
qualitative description of the physics of the low-energy degrees of freedom.
This approach has revealed the existance of a deep conection between the
phenomenon of superfluidity and the FQHE.

Although a lot of progress has been made in the understanding of the
FQHE, the electromagnetic response functions for a generic
incompressible states have not been calculated and, similarly, the
spectrum of collective excitations has not been determined for a
general state.
In a previous work \refto{lopez1}, we presented a theory of the FQHE based on a
second-quantized fermion path-integral approach. There we showed that the
problem of interacting electrons moving on a plane in the presence of an
external magnetic field is equivalent to a family of systems of fermions bound
to an even number of fluxes described by a Chern-Simons gauge field. The
semiclassical approximation of this system has solutions that describe
incompressible-liquid states, Wigner crystals, and solitonlike defects. We
studied the gaussian fluctuations around the liquid-like solution.

In this paper we use the fermion field theory  of reference
[\cite{lopez1}] to
study the collective excitations of the fully polarized FQHE states in
the sequence $1/\nu=1/p+2s$ (where $p$ and $s$ are two positive
integers) recently introduced by Jain\refto{jain1}. In 1986 Girvin,
MacDonald and Platzman\refto{girvin2} used the single mode approximation
to obtain the lowest collective mode in the lowest Landau level (the
{\it intra} mode) for the states in the Laughlin
sequence\refto{laughlin1}
($p=1$ in our notation). Our results include, in addition to the
{\it intra} mode, the {\it inter} mode. In fact we find that, in general
there is a rich spectrum of collective modes. We also find that there are
two modes converging to the cyclotron frequency, and that in general,
these two modes have different spectral weights in the
density correlation function. We discuss in detail the form of the
dispersion curves and the spectral
weights of the various modes for different types of pair interactions.
It should be possible to observe these modes in resonant Raman
scattering
experiments. Recently, the magnetoplasmon modes of integer Hall states
have been observed in light scattering experiments\refto{pinczuk}.

The semiclassical approach of reference [\cite{lopez1}] is
closely related in spirit to the theories of anyon superfluidity. In
both cases an argument is given by which a system of interacting
particles (electrons in the case of the FQHE) is seen to be exactly
equivalent to a system of fermions coupled to a Chern-Simons gauge field
with a properly chosen coupling constant. The mean field theory then
strips the fluxes from the fermions, to which they are locally bound,
and
replaces the fluxes by an average. While this approximation is certainly
very appealing, it has the serious problem that it breaks a number of
space-time symmetries quite explicitly. In particular, it breaks both
Galilean and Magnetic invariance. It turns out that the leading quantum
fluctuations around this state, \ie\ the collective modes, restore
 these symmetries, in the uniform ${\vec Q} \to 0$ limit, already
at the gaussian level. Indeed,
we find that the quadratic or gaussian level of the semiclassical
expansion
gives the correct value of the Hall conductance of the system. Also, at
this level, we verify that the leading
order of the density correlation function saturates the {\it f}-sum rule.
This is an essential result to show that the absolute value squared of the
wave function of all the (incompressible) liquid states has the Laughlin
form at very long distances, in the thermodynamic limit \refto{lopez2}.
As an application of our results, we derive the form of the response to
external test charges and fluxes. Using the same methods, we study the
problem of an interacting
gas of anyons, and we find the spectrum of collective excitations.

The paper is organized as follows. In Section II we review the fermion
Chern-Simons theory
for the FQHE developed in reference \cite{lopez1}. We discuss the problem of
the violation of Galilean invariance by the mean field solution, and its
restoration after the fluctuations are considered at the gaussian level.
In Section III we calculate explicitly the electromagnetic response
functions, discuss their analytic properties, the spectrum of collective
excitations
and some experimentally observable consequences of our results. We also
calculate the Hall response of the system, and verify the saturation of the
{\it f}-sum rule. In Section IV we study the response of the system to
external charges and fluxes. We give explict expressions for the
asymptotic long distance form of the induced charge and current density
profiles, for different types of pair interactions. In Section V we
apply our methods to the problem of the interacting anyon gas, and derive their
spectrum of collective modes. Section VI is devoted to the conclusions.

\head{II.Review of the Chern-Simons field theory for the FQHE}

In this section we review the Chern-Simons field theory for the FQHE that we
developed in reference [\cite{lopez1}].
Our work was motivated by the following argument due to Jain
\refto{jain1}.  The Laughlin wave function,
$$
\Psi (z_1, \dots ,z_N)=\prod_{i<j}(z_i-z_j)^m \;\; \exp(-\sum_{i=1}^N {\vert
z_i \vert^2 \over 4 l^2}).\eqno(2.1)
$$
can be factorized as follows
$$
\Psi(z_1, \dots, z_N)=\prod_{i<j}(z_i-z_j)^{m-1}   \;\;
\chi_1(z_1,\dots,z_N).\eqno(2.2)
$$
where $\chi_1$ is the wave function for a completely filled lowest Landau
level
$$
\chi_1(z_1,\dots,z_N)=\prod_{i<j}(z_i-z_j) \;\; \exp(-\sum_{i=1}^N {\vert
z_i \vert^2 \over 4 l^2}).\eqno(2.3)
$$
Here the set of labels $\{z_i\}$ ($i=1, \dots, N$) are the coordinates
of the $N$
electrons in complex notation ($z=x+iy$) and $l$ is the cyclotron radius.
The {\it odd} integer $m$ is equal to the inverse of the filling fraction
$\nu= {N\over N_{\phi}}=\frac{1}{m}$ of the lowest
Landau level, where $N_{\phi}$ is the total number of flux quanta going
through a sample of linear size $L$, $N_{\phi}={1\over 2\pi}[BL^2/(\hbar
c/e)]$.
In analogy with Laughlin's construction of the quasihole wavefunctions,
Jain observed that the
phases associated with the first factor in eq~(2.2) can be
thought to represent an even
number ($m-1$) of fluxes which are attached to each coordinate $z_i$
where an
electron is present. Since the electrons bind to an even number of flux quanta,
they retain their fermion character.

This observation suggests the possibility of studying the FQHE as an
Integer Quantum Hall
Effect (IQHE) of bound states, \ie\ composites of electrons attached to
an even
number of fluxes, filling up an integer number of Landau levels of the
unscreened
part of the field. In order to do so, we need a theory where particles and
fluxes are bound together. That is precisely what the Chern-Simons gauge
theory does. In 1982
Wilczek \refto{wilczek1} pointed out that a particle current coupled to
a
Chern-Simons (CS) gauge field produces states with fractional statistics
through the binding of particles to fluxes. Therefore, if we want to get the
Laughlin wave function by attaching $m-1$ fluxes to each electron, it is
reasonable to think that the theory should contain fermions coupled to a
Chern-Simons gauge field with an appropriate value of the Chern-Simons coupling
constant $\theta$.

Following these ideas, in reference [\cite{lopez1}] we studied the
problem of a system
of interacting electrons moving on a plane in the presence of an external
uniform magnetic field $B$ perpendicular to it.
In the second quantized language, the action for this system is given by
$$
\eqalign{{\cS}=\int & d^3 z \; \left\{ \psi^*(z) [i D_0 +\mu]\psi(z) -
{1 \over2 m}|{\vec D}\psi(z)|^2 \right\} \cr - &
\12 \int  d^3z \int d^3z'\;(|\psi(z)|^2-{\bar\rho}) V(|{\vec z}-{\vec z}'|)
(|\psi(z')|^2-{\bar\rho}) \cr}          \eqno(2.4)
$$
where ${\bar \rho}$ is the average particle density, $\psi(z)$ is a second
quantized Fermi field, $\mu$ is the chemical potential and $D_{\mu}$ is the
covariant derivative which couples the fermions to the external electromagnetic
field $A_{\mu}$ . The electrons are assumed to have an interparticle
interaction governed by the pair potential $V(|\vec r|)$. In what follows we
will assume that the pair potential has either the Coulomb form, \ie\
$V(|\vec r|) = {q^2 \over |\vec r|}$, or that it represents a short range
interaction such that in momentum space it satisfies that ${\tilde V}(\vec Q)
{\vec Q}^2$ vanishes at zero momentum. This includes the case of
ultralocal potentials (\ie\ with a range smaller or of the same order as
the cyclotron length $l$), in which case we can set ${\tilde V}(0) = 0$,
or short range potentials with a range longer than $l$ such as a Yukawa
interaction.

In reference [\cite{lopez1}] we showed that this system is equivalent to a
system of interacting electrons coupled to an additional statistical
vector potential
$a_{\mu}$ ($\mu=0,1,2$) whose dynamics is governed by the Chern-Simons
action
$$
{\cal S}_{\rm cs}=\int d^3x\;{\theta\over 4}\epsilon_{\mu \nu \lambda}
{a}^{\mu}
{\cal F}^{\nu \lambda}     \eqno(2.5)
$$
provided that the CS coupling constant satisfies $\theta={1\over 2
\pi}{1\over 2s}$,
where $s$  is an arbitrary integer. In eq~(2.5) $x_0,x_1 \;{\rm and}\; x_2$
represent the time and the space coordinates of the electrons respectively,
and ${\cal F}^{\nu \lambda}$ is the field tensor for the statistical gauge
field, ${\cal F}^{\nu \lambda}=\partial^{\nu} {a}^{\lambda}-\partial^{\lambda}
{a}^{\nu}$. In the equivalent theory the covariant derivative given by
$$
D_{\mu}=\partial_{\mu}+i{e\over c}A_{\mu}+i a_{\mu}.\eqno(2.6)
$$
couples the fermions to the statistical gauge field and to the external
electromagnetic field.
For arbitrary values of $\theta$, the system is a set of {\it anyons} with
statistical angle $\delta= {1\over 2 \theta}$, measured with respect to Fermi
statistics. On the other hand, if $\theta={1\over 2 \pi}{1\over 2s}$,
then $\delta=2 \pi s$ and the system still represents fermions.

The Chern-Simons action implies a constraint for the particle density
$j_0({\vec x})$ and the statistical flux $\cal B$, given by
$j_0({\vec x})=\theta {\cal B}(\vec x)$.
This relation states that the electrons coupled to a statistical
gauge field with Chern-Simons coupling constant $\theta$ see a
statistical flux per particle of
${1\over \theta}$. Hence, for $\theta={1\over 2 \pi} {1\over 2s}$, each
fermion picks up a statistical flux equal to ${1\over \theta}=2 \pi (2s)$, \ie
an {\it even} number of flux quanta ($2s$) is attached to each particle.
Hence, if the coefficient of the Chern-Simons
term is chosen in such a way that an even number of flux quanta get attached
to each electron, all the physical amplitudes calculated in this theory are
identical to the amplitudes calculated in the standard theory, in which the
Chern-Simons field is absent. Of course, this is true provided that the
dynamics of the statistical gauge fields is taken into account exactly.

In the scheme that we presented in reference [\cite{lopez1}], the dynamics of
the Chern-Simons gauge fields is taken into account in a semiclassical
expansion, which is a sequence of well controlled approximations.
In practice , we consider the leading and next-to-leading order in the
semiclassical approximation.
Using the constraint enforced by the Chern-Simons term, the action
becomes (in units in which $e=c=\hbar=1$)
$$
\eqalign{{\cS}_{\theta}=\int & d^3 z \; \left\{ \psi^*(z) [i D_0 +\mu]\psi(z) -
{1 \over2 m}|{\vec D}\psi(z)|^2 + {\theta\over 4}\epsilon_{\mu \nu \lambda}
{a}^{\mu} {\cal F}^{\nu \lambda}   \right\} \cr - &
\12 \int  d^3z \int  d^3z'\;(\theta {\cal B} (z)-{\bar\rho})
V(|{\vec z}-{\vec z}'|) (\theta {\cal B}(z') -{\bar\rho}).\cr}   \eqno(2.7)
$$
The quantum partition function for this problem is , at zero temperature
$$
\cZ=\int \cD \psi^* \cD \psi \cD a_{\mu} \exp (i S_{\theta}).\eqno(2.8)
$$
Since the action is quadratic in the fermions, they can be integrated out.
The effective action (${\cal S}_{eff}$) is given by the sum of the fermion
contribution to the effective action (the logarithm of the fermion
determinant, which represents the fermionic fluctuations), the
Chern-Simons
term, and the interaction term. The resulting theory can be treated
within
the saddle-point-expansion (SPA). The external electromagnetic field can
be written as a sum of two terms, one representing the uniform magnetic
field $B$, and a small fluctuating term ${\tilde A}_{\mu}$ whose average
vanishes everywhere. The latter will be used to probe the electromagnetic
response of the system.

The path integral $\cZ$ can be approximated by expanding its degrees of freedom
in powers of the fluctuations, around stationary configurations of
${\cal S}_{eff}$ .
This requirement
yields the classical equations of motion. These equations have many possible
solutions, \ie\ fluid states, Wigner crystals and non-uniform states
with vortex-like configurations. We will only consider
solutions with uniform particle density, \ie the liquid phase solution.
This is the {\it average field approximation} (AFA), which can be
regarded as a mean field approximation. At the mean
field level the electrons see a total flux $B_{\rm eff}$, equal to the
external magnetic flux partially screened by the average Chern-Simons flux,
\ie $B_{\rm eff}=B+\langle \cB\rangle=B-{{\bar\rho}\over \theta}$.
It is easy to see that the uniform saddle-point state has a gap only if the
effective field $B_{\rm eff}$ is such that the fermions fill exactly an
integer number $p$ of the effective Landau levels, \ie those defined by
$B_{\rm eff}$.
In other words, the AFA to this theory yields a state with an energy gap if
the filling fraction satisfies ${1\over \nu}= {1\over p}+2s$, where $p$ and
$s$ are arbitrary positive integers. Only for incompressible states
(\ie with a gap) the perturbative expansion is meaningful. If the system is
compressible, i.e, gapless, the perturbative expansion breaks down. The
breakdown is signalled by infrared divergencies at low temperatures.
This is what happens if an effective Landau level is not completely filled.
The case of a half-filled effective Landau level was analyzed recently by
Halperin, Lee and Read\refto{dreamteam}.

Thus, at the mean field level, the FQHE of a
gas of fermions in a uniform magnetic field is equivalent to an IQHE for
fermions bound to an even number of flux quanta in the presence of a
partially screened external magnetic field.
Now we consider the gaussian (or semiclassical) fluctuations
of the statistical vector potential ${\tilde a}_{\mu}$ around the mean-field
state. Unlike other mean field approaches (such as Hartree-Fock), the
gaussian corrections {\it must} alter the {\it qualitative} properties
of the state described by the AFA. The reason is that the AFA violates
explicitly space-time symmetries, such as Galilean invariance (more
generally, {\it magnetic invariance}) which, for translationally
invariant
systems, must remain unbroken and unchanged. Thus the center of mass of
the system must execute a cyclotron-like motion at, exactly, the
cyclotron
frequency of non interacting electrons in the full external magnetic
field, as demanded by Kohn's theorem\refto{kohn}. A na{\"\i}ve
application of the AFA would suggest that the cyclotron frequency is
renormalized downwards since the effective field seen by the composite
fermions is smaller than the external field $B$. Hence, the {\it
magnetic algebra} may appear to have changed. We will see below that the
gaussian fluctuations yield the correct cyclotron frequency and,  thus,
restore the correct magnetic algebra.

We will now review the the semiclassical expansion for this
system\refto{lopez1}. At the gaussian level, the effective action
for ${\tilde
a}_{\mu}$ is
$$
\eqalign{S^{(2)}({\tilde a}^{\mu},{\tilde A}^{\mu})
                            =&\12 \int d^3x \; d^3y\; {\tilde a}^{\mu}(x)\;
             \Pi^{(0)}_{\mu\nu}(x,y) \; {\tilde a}^{\nu}(y) +
          {\th \over 4} S_{\rm cs}({\tilde a}_{\mu}-{\tilde A}_{\mu})\cr
 -&{\theta^2 \over 2} \int d^3 x d^3 y \;  \left[ {\tilde \cB}(x)-
{\tilde B} (x) \right] \; V(x-y) \;
 \left[ {\tilde \cB}(y)- {\tilde B} (y) \right] \cr} \eqno(2.9)
$$
Eq~(2.9) holds provided that the non-quadratic dependence in the
fluctuating part of the statistical vector potentials ${\tilde a}_\mu$
is small. Recall that these non-quadratic terms results from expanding the
(logarithm) of the fermion determinant in powers of the fluctuations
around the average field approximation. The kernels that enter in the
expressions for these terms are (connected) current correlation
functions (or response functions) of the mean field theory. Thus,
the tensor $\Pi^{(0)}_{\munu}$ is the {\it polarization tensor} of the
equivalent fermion problem at the mean field level, and it is obtained
by expanding the fermion determinant up to quadratic order in the statistical
gauge field. It was shown in reference
[\cite{lopez1}] that this tensor is transverse (as a result of gauge
invariance), analytic in ${\vec Q}^2\over {B_{\rm eff}}$ and that it has simple
poles at  $\omega=k {\bar \omega_c}$ (with $k$ an integer),
where
${\bar \omega_c}\equiv \omega_c/(2sp+1)$ is the cyclotron frequency
associated with the effective magnetic field $B_{\rm eff}$. As a
result, $\Pi^{(0)}_{\munu}$ has a gradient expansion in powers of the inverse
of the effective magnetic field $1\over {B_{\rm eff}}$, or equivalently, in
powers of the inverse of the external magnetic field $1\over B$. In fact, the
dimensionless parameter of this expansion is ${\vec Q}^2\over B$ (we are
working in a system of units such that $\hbar = c =e=1$). It also turns out
that, within this approximation, the limits of $B \rightarrow \infty$ and
$M \rightarrow 0$ are not equivalent (see the explicit form of
$\Pi^{(0)}_{\munu}$ given in Appendix B of reference [\cite{lopez1}]).

The non-quadratic terms in ${\tilde a}_\mu$ in the effective action are of
the form
$$
S_{\rm eff}=S^{(2)}({\tilde a}^{\mu},{\tilde A}^{\mu})+\frac{1}{3!}\int
d^3x_1\; d^3x_2\; d^3x_3\;{\tilde a}^{\mu}(x_1)\;
{\tilde a}^{\nu}(x_2)\; {\tilde a}^{\lambda}(x_3)\;
\Pi^{(0)}_{\mu\nu\lambda}(x_1,x_2,x_3) \;+\ldots
\eqno(2.10)
$$
where the kernel $\Pi^{(0)}_{\mu\nu\lambda}(x_1,x_2,x_3)$ represents a
three-point current correlation function in the mean field theory. Thus,
in the language of Feynman diagrams, while $\Pi^{(0)}_{\mu\nu}(x_1,x_2)$
can be viewed as a fermion bubble with two amputated external collective
mode lines, $\Pi^{(0)}_{\mu\nu\lambda}(x_1,x_2,x_3)$ again has one
fermion loop tied to three amputated external collective mode lines
${\tilde a}_\mu$. Each one of these non-quadratic kernels have the same
gauge invariance (\ie\ transversality) and analytic properties as the
gaussian (or RPA) kernel. In particular, this means that, in momentum
and frequency space, these kernels must be a linear combination of
tensors (of the appropiate rank) which have the correct transversality
properties, times a set of functions which are analytic in ${\vec Q}^2$ and
have
poles at frequencies equal to an integer multiple of the effective
cyclotron frequency. Therefore, the non-quadratic terms necessarily have powers
of ${\vec Q}^2 \over {B_{\rm eff}}$ (for each one of the external momenta and
frequency entering the fermion loop) which are higher than the ones
found at the quadratic level. Since the mean field theory has an integer
number of filled Landau levels, the energy denominators of the kernels do not
change this counting in powers of ${\vec Q}^2 \over {B_{\rm eff}}$.
In conclusion, the expansion of the fermion determinant, and hence of the
effective action, is actually an expansion in powers of
${\vec Q}^2\over B_{\rm eff}$, or equivalently, in powers of
${\vec Q}^2\over B$. However, an expansion in
powers of ${\vec Q}^2\over B$ is also a gradient expansion. Thus, the
gradient expansion and the semiclassical expansion mix and are not
independent from each other.

The semiclassical expansion is obtained according to the following
rules. The propagator for the fluctuations, which represent collective
modes, is the inverse of the kernel of the gaussian action.
Since the pair
potential enters only through the propagator for the fluctuations,
the perturbation theory is not
an expansion in the powers of the pair interaction.
{}From
this point of view, this expansion is very different from conventional
expansions around the Hartree  and Hartree-Fock approximations. The
vertices of the expansion are the kernels for the non-quadratic terms.
This expansion lacks a natural small parameter (\ie\ a coupling
constant) and it should be regarded, like all semiclassical expansions,
as an expansion in the number
of fermion loops (\ie\ RPA plus corrections). One should keep in mind,
however, our previous discussion on its exactness in powers of
${{\vec Q}^2 \over B}$. In what follows
we will make extensive use of the formal properties of this expansion.

The semiclassical expansion has many features in common with the
perturbation theory that Fetter, Hanna and Laughlin (FHL)\refto{fhl}
have
developed for the treatment of the anyon gas. Although, superficially,
they look very different, it is easy to check that it is in fact the
same procedure. The starting point, in both cases, is the average field
approximation. However, when FHL study the fluctuations they fix the
Coulomb gauge ${\vec \bigtriangledown} \cdot {\vec a}=0$. In this gauge,
and by making use of the Chern-Simons constraint $\rho(x)=-\theta
\cB(x)$, one can eliminate the gauge field altogether and the resulting
hamiltonian contains non-local three and four fermion interactions. In
FHL, these non-quadratic terms are dealt within a Hartree and Hartree-Fock
with various improvements adopted to keep track of gauge invariance.
Our functional methods keep track of gauge invariance automatically. The
expansions look different only because of the different choice of gauge.
Our propagators at the gaussian (or semiclassical level) are the
Random Phase Approximation (RPA) propagators (although in a different
gauge).
While being equivalent, our approach is conceptually simpler
and the computations are more direct.

Up to this point we have reviewed the main features of the theory introduced
in reference [\cite{lopez1}]. In the next section we will use the effective
action of eq~(2.9) to calculate the full electromagnetic response
functions of this theory.

\head{III. Electromagnetic response functions for the FQHE}

Since the effective action $S^{(2)}$, eq~(2.9), is quadratic
in ${\tilde a}_{\mu}$, we can integrate out this
field and obtain the effective action for the electromagnetic fluctuations
${\tilde A}_{\mu}$, $S_{\rm eff}^{\rm em}({\tilde A}_{\mu})$ . We will
use this effective action to calculate the full electromagnatic response
functions at the gaussian level. Since this calculation is based on a
one loop effective action for the fermions (\ie\ a sum of fermion
bubble diagrams), this approximation amounts
to a random phase correction to the average field approximation.

In order to integrate out the statistical gauge
field ${\tilde a}_{\mu}$ we must fix the gauge. The
electromagnetic effective action, being gauge invariant, is
independent of the choice of gauge for the statistical gauge fields in
the path integral. We fix the gauge $\partial _{\mu} {\tilde a}^{\mu}={\alpha}$
 (where $\alpha$ is an arbitrary real number)  using the standard
Faddeev-Popov
procedure (see footnote [\cite{ramond}]). The result is explicitely
gauge invariant and
all dependence on the parameter $\alpha$ cancels out. At the one-loop
level (governed by the effective action of eq~(2.9)) we need to know the
inverse of the polarization tensor
of the equivalent fermion problem, $\Pi^{(0)}_{\munu}$. In reference
[\cite{lopez1}], we
showed
that $\Pi^{(0)}_{\munu}$ can be written in terms of three gauge invariant
tensors, an ${\vec E}^2$ term, a ${\vec B}^2$ term, and a Chern-Simons term.
These three tensors plus  $ B {\vec{\nabla}}.{\vec E}$ and a gauge
fixing term (such as ${1\over 2\alpha}({\partial _{\mu}}{{\tilde a}^{\mu}})^2$
which corresponds to the Landau-Lorentz gauge if $\alpha \rightarrow 0$)
close an algebra that
can be used to invert the polarization tensor and to calculate explicitly the
electromagnetic response functions.

After integrating out the statistical gauge field in eq~(2.9), the effective
action for the electromagnetic fluctuations ${\tilde A}_{\mu}$ turns out to be
$$
{\cal S}_{\rm eff}^{\rm em} ({\tilde A}_{\mu}) = {1\over 2}
                  \int d^{3}x \int d^{3}y {\tilde A}_{\mu}(x) K^{\mu \nu}(x,y)
                   {\tilde A}_{\nu} (y)              \eqno(3.1)
$$
Here $K^{\mu\nu}$ is the electromagnetic polarization tensor. It measures the
linear response of the system to a weak electromagnetic perturbation.
Its components can be written in momentum space as follows
$$
\eqalign{ K_{00} &= {\vec Q}^2  K_{0}(\omega, {\vec Q}) \cr
          K_{0j} &= {\omega} Q_{j} K_{0}(\omega, {\vec Q})
                  + i {\epsilon _{jk}}Q_{k} K_{1}(\omega, {\vec Q}) \cr
          K_{j0} &= {\omega} Q_{j} K_{0}(\omega, {\vec Q})
                  - i {\epsilon _{jk}} Q_{k} K_{1}(\omega, {\vec Q}) \cr
          K_{ij} &= {\omega}^2 {\delta _{ij}} K_{0}(\omega, {\vec Q})
                  - i {\epsilon _{ij}} {\omega} K_{1}(\omega, {\vec Q})
           + ({\vec Q}^2 {\delta _{ij}}- {Q_i}{Q_j})K_{2}(\omega, {\vec Q})\cr}
            \eqno(3.2)
$$
where the functions $K_{l}(\omega, {\vec Q})$ ($l=0,1,2$) are given by
$$
\eqalignno{K_{0}(\omega, {\vec Q}) &= - {\theta ^2} \;
                           { {\Pi _{0}} \over
                            D(\omega,{\vec Q})} &(3.3)\cr
           K_{1}(\omega, {\vec Q}) &=\theta
                  +{\theta}^2 \; {(\theta +{\Pi _1} )
                             \over D(\omega,{\vec Q})}
     +{\theta}^3 {\tilde V}(\vec Q) {\vec Q}^2 \; {{\Pi_{0}}
                             \over  D(\omega,{\vec Q})}   &(3.4)  \cr
           K_{2}(\omega, {\vec Q}) &=-{\theta}^2 \;
   {{\Pi_2} +{\tilde V}(\vec Q)\; ({\omega}^2 \;
             {\Pi_0}^2 -{\Pi_1}^2 +{\vec Q}^2\; {\Pi_0} \;
    {\Pi_2})\over D(\omega,{\vec Q})}   &(3.5) \cr}
$$
and
$$
D(\omega,{\vec Q}) = {\Pi_0}^2{\omega}^2-({\Pi_1}+\theta)^2+{\Pi_0}
                     ({\Pi_2}-{\theta}^2{\tilde V}(\vec Q)){\vec Q}^2
\eqno(3.6)
$$
The coefficients $\Pi_{l}$ ($l=0,1,2$) are functions of $\omega$ and
${\vec Q}$, and are given explicitly in Appendix B of reference
[\cite{lopez1}].
${\tilde V}(\vec Q)$ is the Fourier transform of the interparticle pair
potential.
As we mentioned before, we needed to include a gauge fixing term to be able to
compute the functional integral in eq~(2.9). But at the end of the calculation
all the terms which contain the gauge fixing coefficient ($\alpha$) cancel
each other and the final result for the response functions is, as it must be,
gauge invariant.
The other tensor that we have introduced to make the calculations,
$B {\vec{\nabla}}.{\vec E}$, is not present in the final answer either.

We want to stress here that the thermodynamic limit is crucial for the
accuracy of our results. Notice first that in the electromagnetic
effective action of eq~(3.1) we are neglecting higher order response
functions, \ie correlation functions of three or more currents or densities.
We have shown in Section II that these higher order correlation functions
have higher order powers of ${{\vec Q}^2 \over B}$ than the quadratic
term. Strictly speaking, these terms are not neglectible for a finite system
because, in this case, there is a minimum value that the momentum can
take,
determined by the linear size of the system $L$, \ie\
$|\vec Q| > {1\over L}$. But in the thermodynamic limit,
$L \rightarrow \infty$ and the momentum can go to zero. In other words,
only for an infinite system one is allowed to keep only the quadratic term in
the electromagnetic action, eq~(3.1), and to neglect the higher order
correlation functions.

The electromagnetic response functions determined by $K_{\mu\nu}$ have
the following properties:
\item{i)} We saw in reference [\cite{lopez1}] that the polarization tensor
at mean field
level, $\Pi^{(0)}_{\munu}$, has poles at every value of the effective
cyclotron
frequency (${\bar \omega}_{c}\equiv {B_{\rm eff} \over M}$). This corresponds
to the physical picture, at mean field level, of an IQHE of the bound states
in the presence of a partially screened external magnetic field,
$B_{\rm {eff}}$. Once we take into account the Gaussian fluctuations,
it is easy to prove that all this poles that are present in the numerator and
the denominator of the $K_{\munu}$ components through the $\Pi_{j}$'s, cancel
out, and the poles of the response functions are determined only by the zeroes
of their denominator, $D(\omega, \vec Q)$. In other words, the collective
excitations of this system will be determined only by the zeroes of
$D(\omega, \vec Q)$.
\item{ii)} The leading order term in ${\vec Q}^2$ of the $K_{00}$
component of the polarization tensor saturates the $f$-sum rule.
\item{iii)} The gaussian fluctuations of the statistical gauge field are
responsible
for the FQHE. In particular, the gaussian corrections yield the exact
value for the Hall conductance.

In the remaining of this section we will discuss these properties and
their experimentally accessible consequences in detail.

\item{a)} The Spectrum of Collective Excitations:

For simplicity, we have studied the zeroes of $D(\omega, \vec Q)$ in two cases,
 when the number of effective Landau levels filled is $p=1$ and $p=2$.
The (more tedious) case of general $p$ can be studied by straightforward
application of the same methods.
We have looked for solutions of the form
$\omega ^2 = (k {\bar \omega}_c )^2 + \beta \;
({{\vec Q}^2\over 2 {B_{\rm eff}}})^{\gamma}$,
 where $\beta$ and $\gamma$ are two constants to be determined.
Thus, we substitute this expression into the functions
$\Pi _i (\omega , {\vec Q})$ which appear in $D(\omega , {\vec Q})$, and
expand both the numerators and the denominators in powers of $\vec Q$.
Looking
at the coefficients of the leading and subleading terms of this
expansion, we are able to determine the values of $\beta$ and $\gamma$
for all the proposed solutions. This procedure is quite straightforward
to carry out. Only the modes with $k=1,m$ require special care.

\noindent{Case $p=1$}

In this case the filling fraction is $\nu = {1\over m}$, where $m=1+2s$, \ie\
the Laughlin sequence.

We find that there is a family of collective modes whose zero-momentum
gap is $k {\bar \omega}_{c}$, where $k$ is an integer number different from
$1$ and $m$, and whose dispersion curve $\omega_{k} ({\vec Q})$ is
$$\omega_{k} ({\vec Q})= \Big [ (k {\bar \omega}_{c})^2 +
                         ({{\vec Q}^2\over 2 {B_{\rm eff}}})^{k-1}\;
                         {\bar \omega}_{c}^2 \;
            {2k (m-1)(k-1) \over {(k-1)!\; (k-m)}} \Big ] ^{1\over2} \eqno(3.7)
$$
The residue in $K_{00}$ corresponding to this pole is
$$
Res(K_{00},\omega _{k}({\vec Q})) = -{\vec Q}^2 \;
                                   { \omega}_{c}{\nu \over 2\pi}\;
                                ({{\vec Q}^2\over 2 {B_{\rm eff}}})^{k-1}
                               {2k (m-1)(k-1)\over {(k-1)!(k-m)(k^{2}-m^{2})}}
                                  \eqno(3.8)
$$
The cases $k=1,m$ have to be treated separately. In general, we find
that there is no mode with a  zero momentum gap at ${\bar \omega}_{c}$.
Instead, at ${\vec Q}=0$, there is a doubly degenerate mode with a gap
at
$ \omega_{c}$. This degenerate cyclotron mode can be viewed as the
mixing of the modes with $k=1$ and with $k=m$.
Thus, the mode
with $k=1$ has been ``pushed up" to the cyclotron frequency (at ${\vec
Q}=0$). Halperin
\etal\ \refto{dreamteam} have recently found a similar result.
For ${\vec Q} \not=0$, the degeneracy is lifted and these two modes have
different dispersion curves.

For the special case of $\nu = {1\over 3}$, \ie\ $m=3$, this effect is
particularly important. The dispersion relations for the cyclotron modes
are given by
$$
\omega_{\pm} ({\vec Q}) = \Big [ {\omega_{c}}^2 +
            ({{\vec Q}^2\over 2 {B_{\rm eff}}}){{\bar \omega}_{c}^2 \over 2}
            \alpha _{\pm }\Big ] ^{1\over 2}  \eqno(3.9)
$$
where
$$
\alpha _{\pm} = 8+ {2M {\tilde V}({0}) \over 2\pi} \pm
 {\Big (} (8+ {2M {\tilde V}(0) \over 2\pi})^2 + 288 {\Big )}^{1\over 2}
             \eqno(3.10)
$$
The residues corresponding to these poles are
$$
Res(K_{00},\omega _{\pm }({\vec Q})) = -{\vec Q}^2 \;
                                { \omega}_{c}{\nu \over 2\pi}\;
                                ({1+ {288\over \alpha_{\pm}^2}})^{-1}
                                  \eqno(3.11)
$$
For  $\nu = {1\over m}$, $m \geq 5$, the collective modes whose
zero-momentum gap is the cyclotron frequency, $\omega _c$, are
$$
 {\omega_{+}}(\vec Q) = \Big [{\omega _{c}}^2 +
                          ({{\vec Q}^2\over 2 {B_{\rm eff}}}) \;
                         {\bar \omega}_{c}^2 \; {\Big(} 4 {{(m-1)}\over (m-2)}+
        {2M {\tilde V}(0) \over 2\pi}{\Big)}\Big ] ^{1\over2}\eqno(3.12)
$$
with residue
$$
Res(K_{00},\omega _{+}({\vec Q})) = -{\vec Q}^2
                                {\omega}_{c}{\nu \over 2\pi}  \eqno(3.13)
$$
The other cyclotron mode has the dispersion
$$
{\omega_{-}}(\vec Q) = \Big[ {\omega _{c}}^2 -
                        ({{\vec Q}^2\over 2 {B_{\rm eff}}})^{m-2}\;
                        {\bar \omega}_{c}^2 \;
                        {{4 m^2 (m-1)^2} \over {(m-1)!}}\;
                        \Big( 4 {{(m-1)}\over (m-2)}+
                        {2M {\tilde V}(0) \over 2\pi}\Big)^{-1}
                        \Big] ^{1\over2}       \eqno(3.14)
$$
with residue
$$
Res(K_{00},\omega _{-}({\vec Q})) = -{\vec Q}^2 \;
                                {\omega}_{c}\; {\nu \over 2\pi}\;
                                ({{\vec Q}^2\over 2 {B_{\rm eff}}})^{m-3} \;
                        {{4 m^2 (m-1)^2} \over {(m-1)!}} \;
                        \Big( 4 {{(m-1)}\over (m-2)}+
                        {2M {\tilde V}(0) \over 2\pi}\Big)^{-2}
                        \eqno(3.15)
$$
The above results are valid only if the pair potential ${\tilde V({\vec Q})}$,
has a gradient expansion in powers of $\vec Q$, \ie\ for short range
interactions.
${\tilde V}(0)$ stands for the leading order term in that expansion.

If the pair potential has the Coulomb form \ie\
${\tilde V}({\vec Q}) = {2\pi{q^2} \over |{\vec Q}|}$ in two spacial
dimensions,
 both, the dispersion relations with zero-momentum gap at the cyclotron
frequency and their residues get modified. The expressions valid in this case
are, for any allowed value of $m$
$$
 {\omega_{+}}(\vec Q) = \Big [{\omega _{c}}^2 +
                          {|{\vec Q}|\over 2 {B_{\rm eff}}} \;
                         {\bar \omega}_{c}^2 \; {2M{q^2}}
                          \Big ] ^{1\over2}\eqno(3.16)
$$
with the same residue given by eq~(3.13), and
$$
{\omega_{-}}(\vec Q) = \Big[ {\omega _{c}}^2 -
                        {|{\vec Q}|^{2m-3}\over ({2 B_{\rm eff}})^{m-2}} \;
                        {\bar \omega}_{c}^2 \;
                        {{4 m^2 (m-1)^2} \over {2M{q^2}(m-1)!}}\;
                        \Big] ^{1\over2}       \eqno(3.17)
$$
with residue
$$
Res(K_{00},\omega_{-}({\vec Q})) = -{\vec Q}^2 \;
                                {\omega}_{c}\; {\nu \over 2\pi}\;
                       {|{\vec Q}|^{2(m-2)}\over {(2{B_{\rm eff}})}^{m-3}} \;
                      {{4 m^2 (m-1)^2} \over {(2M{q^2})^2(m-1)!}} \;
\eqno(3.18)
$$

\noindent{Case $p=2$}

In this case the filling fraction is $\nu = {2\over m}$ where $m=1+4s$.
The same remarks about the pair potential are valid in this case.
If the pair potential has a gradient expansion in powers of $\vec Q$ the
following results hold.

We find that there is a family of collective modes whose zero-momentum
gap is $k {\bar \omega}_{c}$, with $k \not= 1,m$, and whose dispersion
curve $\omega_{k} ({\vec Q})$ is
$$
\omega_{k} ({\vec Q})= \Big[ (k {\bar \omega}_{c})^2 +
                       ({{\vec Q}^2\over 2 {B_{\rm eff}}})^{k-1} \;
                       {\bar \omega}_{c}^2 \;
                       { (m-1)(k-1)k(k+2) \over {(k-m)(k-1)!}}
                        \Big] ^{1\over2} \eqno(3.19)
$$
The residue corresponding to this pole in $K_{00}$ is
$$
Res(K_{00},\omega _{k}({\vec Q})) = -{\vec Q}^2 \;
                                   {\omega}_{c}{\nu \over 2\pi}
                                   ({{\vec Q}^2\over 2 {B_{\rm eff}}})^{k-1}
                      {(k+2)(k+1)k(k-1)^2\over {(k-1)!(k-m)(k^{2}-m^{2})(m+1)}}
                                  \eqno(3.20)
$$
The collective modes whose zero-momentum gap is the cyclotron frequency,
$\omega _c$, are
$$
 {\omega_{+}}(\vec Q) = \Big[ {\omega _{c}}^2 +
                          ({{\vec Q}^2\over 2 {B_{\rm eff}}}) \;
                    {\bar \omega}_{c}^2 \; 2 \; \Big( 4 {{(m-1)}\over (m-2)}+
            {2M {\tilde V}(0) \over 2\pi}\Big) \Big]^{1\over2}\eqno(3.21)
$$
with residue
$$
Res(K_{00},\omega _{+}({\vec Q})) = -{\vec Q}^2 \;
                                { \omega}_{c}\; {\nu \over 2\pi}  \eqno(3.22)
$$
and
$$
{\omega_{-}}(\vec Q) = \Big[ {\omega _{c}}^2 -
                        ({{\vec Q}^2\over 2 {B_{\rm eff}}})^{m-2} \;
                        {\bar \omega}_{c}^2 \;
                        {{ m^2 (m-1)^2(m+2)} \over {(m-1)!}} \;
                        \Big( 4 {{(m-1)}\over (m-2)} +
                        {2M {\tilde V}(0) \over 2\pi}\Big)^{-1}
                        \Big] ^{1\over2}       \eqno(3.23)
$$
with residue
$$
Res(K_{00},\omega _{-}({\vec Q})) = -{\vec Q}^2 \;
                                {\omega}_{c}\; {\nu \over 2\pi}\;
                                ({{\vec Q}^2\over 2 {B_{\rm eff}}})^{m-3} \;
                        {{m^2 (m-1)^2(m+2)} \over {2(m-1)!}}\;
                        \Big( 4 {{(m-1)}\over (m-2)}+
                        {2M {\tilde V}(0) \over 2\pi}\Big) ^{-2}
                        \eqno(3.24)
$$
If the pair potential has the Coulomb form, the dispersion relations
with zero-momentum gap at the cyclotron frequency become
$$
 {\omega_{+}}(\vec Q) = \Big[ {\omega _{c}}^2 +
                          {|{\vec Q}|\over 2 {B_{\rm eff}}} \;
                    {\bar \omega}_{c}^2 \; 4M{q^2} \Big]^{1\over2}\eqno(3.25)
$$
with residue given by eq~(3.22), and
$$
{\omega_{-}}(\vec Q) = \Big[ {\omega _{c}}^2 -
                        {|{\vec Q}|^{2m-3}\over ({2{B_{\rm eff}}})^{m-2}} \;
                        {\bar \omega}_{c}^2 \;
                        {{ m^2 (m-1)^2(m+2)} \over {2M{q^2}(m-1)!}} \;
                        \Big] ^{1\over2}       \eqno(3.26)
$$
with residue
$$
Res(K_{00},\omega _{-}({\vec Q})) = -{\vec Q}^2 \;
                                {\omega}_{c}\; {\nu \over 2\pi}\;
                        {|{\vec Q}|^{2(m-2)}\over ({2{B_{\rm eff}}})^{m-3}} \;
                        {{m^2 (m-1)^2(m+2)} \over {2(2M{q^2})^{2}(m-1)!}}\;
                        \eqno(3.27)
$$
In this section we have found the spectrum of collective excitations for
some values of the filling fraction. Our results are a generalization of
the work of Kallin and Halperin\refto{kallin} who studied the spectrum
of collective modes for the {\it integer} quantum Hall effect
within the RPA. We find a family of collective modes with dispersion
relations
whose zero-momentum gap is $k{\bar \omega}_{c}$, where $k$ is an
integer number
different from $1$ and $m$. When $k=m$, \ie the zero-momentum gap is the
cyclotron frequency, there is a splitting in the dispersion relation for
finite wavevector. One of these solutions, $\omega_{-}$, has
negative slope for small values of $\vec Q$. Therefore, there must be a roton
minimum at some finite value of the wavevector. Since our results are
accurate
only for small $\vec Q$, our dispersion curves do not apply close to
the roton minimum. Nevertheless, this mode is expected to become damped
due to non-quadratic interactions among the collective modes. On the other
hand, the collective mode with lowest energy, which has $k=2$, is stable
(at least for reasonably small wavevectors) and, at small wavectors, it
disperses downwards in energy. This behavior suggests that there should be a
magnetoroton minimum for this mode. This result is consistent with
the work of Girvin \etal\ \refto{girvin2}.

The splitting of the cyclotron mode for $\nu=1/3$ is a little
puzzling. It only happens
for $\nu=1/3$ and for short range interactions. In all other cases,
only the residue for one of the two cyclotron modes is proportional
to ${\vec Q}^2$. Standard lore has it that Kohn's theorem demands that there
should be one and only one mode converging to the cyclotron
frequency as ${\vec Q}^2 \to 0$ with residue proportional to ${\vec Q}^2$.
Zhang has emphasized this point recently\refto{zhang1}.
It is generally assumed that  Kohn's theorem is valid even at non-zero
wavevectors and that it requires the existance of only one mode with
residue proportional to ${\vec Q}^2$ converging to $\omega_c$ . However,
at non-zero wave vectors, these arguments make the unstated
assumption of the analiticity of the current operators on the
wavevectors. While this may well be correct, it is an additional
assumption  and it does deserve closer scrutiny. The results from our
theory
do indeed predict the existance of only one mode {\it at} $\omega_c$
with residue proportional to ${\vec Q}^2$, which is the statement of
Kohn's theorem. And, also, for all filling
fractions and for all pair potentials (except $\nu=1/3$ and short range
interactions) we do find only one mode with residue ${\vec Q}^2$ even at
non-zero wavevectors. The case $\nu=1/3$ and short range interactions
appears to be exceptional in that we find two modes which coalesce at
the cyclotron frequency as ${\vec Q}^2 \to 0$. But both of these modes have
residue proportional to ${\vec Q}^2$ , with different amplitude, and
together they satisfy the sum rule\refto{puzzle}.
While it is possible
that the non-gaussian corrections may change this result since, in a sense,
these are subleading pieces in ${\vec Q}^2$, these
non-gaussian corrections are expected to be very
small at small wavevectors.

We close this section with a few comments on the validity of this spectrum
of collective modes beyond the semiclassical approximation.
Primarily we have to consider the physics at moderately large wavevectors
and the (expected) effects of non-gaussian corrections. At the gaussian
(RPA) level we found a family of collective modes which, for sufficiently
small momentum, are infinitely long lived (\ie\ the response functions have
delta-function sharp poles at their location). These modes represent
charge-neutral bound states. It is in principle possible that,
for $\vec Q$ sufficiently large, these modes should become damped.
The threshold should occur when the energy of the collective mode becomes
equal to the energy necessary to create the lowest available
two-particle state: a quasiparticle-quasihole pair. In the AFA, the energy
of a pair is equal to ${\bar \omega}_c$. Gaussian fluctuations are expected
to renormalize this energy upwards and to give it a
momentum dependence. This is in principle calculable with the methods of
this paper but this result is not available at present time.
Non-gaussian corrections to the RPA are also expected to give a finite
width to (presumably) all the collective modes but the lowest one. This
is so because the corrections to the semiclassical
approximation are due to effective vertices (due to virtual
quasiparticle-quasihole pairs) which couple the various collective modes
and, thus, induce the higher energy modes to decay down into the lower
modes. However, by gauge invariance, these vertices have a momentum dependence
and should vanish as ${\vec Q} \to 0$. Thus, the width of the higher
energy modes goes to zero as ${\vec Q} \to 0$ and these modes only become
sharp at ${\vec Q}=0$. But at ${\vec Q}=0$ the only accessible mode is the
cyclotron mode (the other modes have a vanishingly small spectral
weight). These arguments strongly suggest that the only truly sharp mode,
{\it at} ${\vec Q}=0$, is the cyclotron mode, which is required to be stable
by Kohn's theorem\refto{kohn}. Since the modes with zero momentum gap at
$k{\bar \omega_{c}}$, $k \geq 3$, are not the collective modes with lowest
energy, it is possible that at finite wavevectors they may also decay into
the collective mode with lowest energy ( the mode with
$k=2$, which has a gap at ${\bar \omega}_c$). These issues remain to be
investigated.

\item{b)} Experimental consequences:

In this section we discuss the experimental consequences of
the results that we have just derived.

The density correlation function can be probed by optical absortion and by
Raman-scattering experiments.

In the first case, the optical absortion is
proportional to the imaginary part of the density correlation function.
We predict that there will be absortion peaks at a discrete set of frequencies
which, for ${\vec Q}\to 0$ converge to
$\omega = k {\bar \omega }_{c}$, where $k$ is an integer number greater than
two. Since the spectral weight of these modes vanishes as ${\vec Q}\to
0$, the associated absorption peaks are,  for a strictly
translationally invariant system,  only observable at non-zero momentum.

In the case of the Raman scattering, the geometry must be such that there is a
component of the incident light wavevector in the plane of the sample.
The Raman spectrum, $I(\omega)$, is also proportional to the imaginary part
of the density correlation function \refto{klein}.

We have seen that in the limit $|{\vec Q}|\ll {l}^{-1}$, where $l$ is
the
magnetic length, most of the weight of $K_{00}(\omega,{\vec Q})$ is in one of
the cyclotron modes. The pole in $K_{00}(\omega,{\vec Q})$ for the lowest
excitation frequency, $\omega _k$ with $k=2$, has a residue which is
proportional to $|{\vec Q}|^4$, \ie it is smaller by a factor of
$|{\vec Q}|^2$ than the residue at the highest weighted mode at the cyclotron
frequency.

We have also found that there is a splitting in the cyclotron modes.
If the pair potential has a gradient expansion in $|{\vec Q}|$, \ie short range
interaction, the pole at $\omega_{-}$ (eq ~(3.14) and (3.23)), has a residue
that is smaller by a factor of $|{\vec Q}|^{2(m-3)}$ than the residue of
$\omega _{+}$ (eq ~(3.12) and (3.21)). The relative Raman intensity,
${I(\omega_{+})\over I(\omega_{-})}$, is proportional to
$({{2B_{\rm eff}} \over {\vec Q}^{2}})^{(m-3)}$ which is a big number within
our approximation.
If the filling fraction is $\nu ={1\over 3}$, both modes have the same ${\vec
Q}^2$
dependence in their spectral weight, but the relative intensity is
$\approx 2.5 $ provided that ${\tilde V}(0)=0$.
Except for $\nu={1\over 3}$, the splitting between the two modes at the
cyclotron frequency satisfies, at leading order in $|{\vec Q}|$,
$\triangle \omega ^2 = \omega _{+}^{2} - \omega _{-}^{2}
= \omega _{+}^{2} - \omega _{c}^{2}$, which is proportional to $|{\vec Q}|^2$.
Up to this order, experimentally one should observe one mode dispersing as
$\omega _{+}$ (eq~(3.12) or (3.21)), and the other as $\omega =\omega _{c}$.
For $\nu = {1\over 3}$ the splitting is also proportional to $|{\vec Q}|^2$.
In this case one should observe both modes ($\omega _{+}$ and
$\omega _{-}$, eq~(3.9)), but with different intensities.

If the pair potential has the Coulomb form, the residue of $\omega_{-}$
(eq ~(3.18) or (3.27)) is smaller by a factor of $|{\vec Q}|^{2(m-2)}$
than the residue of $\omega _{+}$ (eq ~(3.16) or (3.25)), and this is valid
for all the values of the filling fraction that we have studied.
The splitting between these two modes satisfies, at leading order in
$|{\vec Q}|$, $\triangle \omega ^2 = \omega _{+}^{2} - \omega _{c}^{2}$,
which is proportional to $|{\vec Q}|$. For $\nu$ different from $1\over 3$,
the relative intensity between the two modes is proportional to
$({{2B_{\rm eff}}/ {\vec Q}^{2}})^{(m-3)}{M{\tilde V}(\vec Q)}$, which is
bigger than $1$ within our approximation. For $\nu = {1\over 3}$,
the relative intensity is proportional to ${M{\tilde V}(\vec Q)}$. This factor
can be written in terms of the magnetic length and the cyclotron energy
as follows $ {{\tilde V}(\vec Q) / l \over \omega_{c}}$. Since
our approximation is only valid in the limit ${1\over |{\vec Q}|}\gg l$,
the
numerator satisfies ${\tilde V}(\vec Q) / l \gg { 2\pi {q^2} \over l}$. The
second
term in this inequality is the Coulomb energy at the magnetic length, which is
typically of the same order of magnitude than the cyclotron energy. Therefore,
${{\tilde V}(\vec Q) / l\over \omega_{c}} \gg { 2\pi {q^2} / l\over
\omega_{c}}
\approx 1$. In other words, the relative intensity for $\nu ={1\over 3}$ is
also bigger than one.

\item{c)}  Saturation of the $f$-sum rule:

We show now that the long wavelength form of $K_{00}$, found at this
semiclassical level, saturates the $f$-sum rule. This result implies
that the non-gaussian corrections do not contribute at very small
momentum. In a separate publication\refto{lopez2} we have used this
result to show that the absolute value squared wave function of all the
(incompressible) liquid states has the Laughlin form at very long
distances, in the thermodynamic limit.

The retarded density and current correlation functions of this theory are, by
definition
$$
D^R_{\mu \nu}(x,y)  = -i \theta (x_{0}-y_{0}) <G| [J_{\mu}(x),J_{\nu}(y)] |G>
   \eqno(3.28)
$$
where $J_{\mu}$ ($\mu =0,1,2$) are the conserved  currents of the theory
defined by eq~(2.4), and $|G>$ is the ground state of the system.
Using this definition and the commutation relations between the currents,
one can derive the $f$-sum rule for the retarded density correlation
function $D^R_{00}$. In units in which $e=c=\hbar =1$, it states that
$$
\int _{-\infty}^{\infty } \; {d\omega \over 2\pi} \; i \omega
       D^R_{00}(\omega,{\vec Q}) = {{\bar \rho} \over M}{\vec Q}^2 \eqno(3.29)
$$
On the other hand, it is easy to show (see for instance reference
[\cite{book}])
that the polarization tensor $K_{\mu\nu}$ and the density and current
correlation functions $D_{\mu\nu}$ satisfy the following identity
$$
K_{\mu\nu}(x,y) = - D_{\mu\nu}(x,y)
          + <{\delta J_{\mu}(x) \over \delta A_{\nu}(y)}>   \eqno(3.30)
$$
{}From eq~(3.2) and ~(3.3) we see that the leading order term in
${\vec Q}^2$ of the zero-zero component of the electromagnetic response is
given by
$$
K_{00} = - {{\bar \rho}\over M}\; {{\vec Q}^2 \over {{\omega }^2 -{\omega
}^2_c}
            + i \epsilon}
     \eqno(3.31)
$$
where we have used that ${{\bar \rho}\over B} = {\nu \over 2\pi}$.

The correlation functions that we derive from the path integral formalism
are time-ordered. Therefore, if we use the relation between time-ordered and
retarded Green's functions, and eq~(3.30) and ~(3.31), we see that the leading
order term of $K_{00}$ saturates the $f$-sum rule, eq ~(3.29).

It is important to remark that the coefficient of the leading order term of
$K_{00}$ can not be renormalized by higher order terms in the gradient
expansion, nor in the semiclassical expansion. In the case
of the gradient expansion, it is clear that higher order terms have higher
order powers of ${\vec Q}^2$, and then, do not modify the leading order term.
In the case of the corrections to $K_{00}$ originating in higher order terms
in the semiclassical expansion, they also come with higher order powers of
${\vec Q}^2$. The reason of that is essentially the gauge invariance of the
system. This implies that the higher order correlation functions must be
transverse in real space, or equivalently they have higher order powers of
${\vec Q}^2$ in momentum space.
Being higher order terms in the ${\vec Q}^2$ expansion they can not change
the leading order term.

As we have already mentioned, these results hold for any model Hamiltonian
for the two-dimensional electron gas (2DEG) with reasonably local interactions,
 \ie with pair interactions that obey ${\vec Q}^2 {\tilde V}(Q) \rightarrow 0$
as ${\vec Q}^2 \rightarrow 0$.

\item{d)} Hall Conductance:

We show now that, already within our approximations, this state does
exhibit the Fractional Hall Effect with the exact value of the Hall
conductance. We have previously shown elsewhere\refto{lopez1,book} that
this is the case using the effective action of the statistical gauge
fields. Here we show that, as expected, the electromagnetic response
functions exhibit the correct FQHE.

In order to do so, we will calculate the Hall conductance of the whole system.
Since we are only
interested in the leading long-distance behavior, it is sufficient to keep only
with those terms in eq~(3.1) which have the smallest number of derivatives, or
in momentum space, the smallest number of powers of $\vec Q$.
Therefore, from eq~(3.1) and ~(3.2), we see that the leading long distance
behavior (\ie small momentum)  of the effective action for the electromagnetic
field is governed by the Chern-Simons term.
In this limit eq~(3.1) turns out to be
$$
S_{\rm eff}^{\rm em} ({\tilde A}_{\mu}) \approx -{i\over 2}
                       \int {d^{2}Q\over (2\pi)^2} \int {d\omega\over (2\pi)}
                       {\tilde A}_{\mu}(-\omega,-{\vec Q})
                       {K_1(\omega,{\vec Q})} {\epsilon _{\mu\nu\lambda}}
                       { Q^{\lambda}}{\tilde A}_{\nu} (\omega, \vec Q)
                        \eqno(3.32)
$$
where $Q^{0}=\omega$ and $Q^{i}=-Q_{i}$ according with the convention that we
have used in reference [\cite{lopez1}].

To  study the Hall response of the system, we will now consider the
limit of small $\omega$ and
small $\vec Q$.
We have checked that in this theory the two limits commute
$$
{\lim _{{\vec Q} \rightarrow 0}}\; {\lim _{\omega \rightarrow 0}}
{ K_1}(\omega,{\vec Q}) ={{\Pi_{1}(0,0)}\over{\theta + \Pi_{1}(0,0)}}
      \eqno(3.33)
$$
$$
{\lim _{\omega \rightarrow 0}}\; {\lim _{{\vec Q} \rightarrow 0}}
{ K_1}(\omega,{\vec Q}) ={{\Pi_{1}(0,0)}\over{\theta + \Pi_{1}(0,0)}}
      \eqno(3.34)
$$
This is a consequence of the incompressibility of the ground state.
Since $\theta = {1\over 2\pi \;2s}$ and $\Pi_{1}(0,0)=
{p\over 2\pi}$
$$
K_{1}\rightarrow {\nu \over 2\pi} \equiv \th_{\rm eff}      \eqno(3.35)
$$
where $\nu$ is the filling fraction.
The electromagnetic current
$J_{\mu}$ induced in the system is obtained by differentiating the
effective action $S_{\rm eff} ({\tilde A}_{\mu})$ with respect
to the electromagnetic vector potential. The current is
$J_{\mu}={\th_{\rm eff}\over 2} \eps_{\munu \la} {\tilde F}^{\nu
\la}$.
Thus, if a weak external electric field ${\tilde E}_j$ is applied, the
induced current is
$J_k=\th_{\rm eff} \eps_{lk} {\tilde E}_l $.
We can then identify the coefficient $\th_{\rm eff}$ with the {\it
actual} Hall conductance of the system $\si_{xy}$ and get
$$
\si_{xy}\equiv \th_{\rm eff}={\nu\over 2 \pi}\eqno(3.36)
$$
which is a {\it fractional} multiple of ${e^2\over h}$ (in units in
which $e=\hbar=1$). Thus, the uniform states exhibit a Fractional
Quantum Hall effect with the correct value of the Hall conductance.

\head{IV. Electromagnetic Response to an external charge
and flux}

In this section we will study the linear response of the system to a static
charge and a static flux.

Consider the case of a static probe of electric charge $\tilde q$, located at
the origin. The electromagnetic vector potential can be written as
$$
\eqalign{{\tilde A}_{0}({\vec x},t) &= {{\tilde q} \over |{\vec x}|} \cr
{\tilde A}_{j}({\vec x},t) & = 0,  \; j=1,2 \cr}            \eqno(4.1)
$$
or in momentum space
$$
{\tilde A}_{0}(\omega, {\vec Q}) =
          {(2\pi)}^2 \delta (\omega)\; {{\tilde q} \over |{\vec Q}|}\eqno(4.2)
$$
The electromagnetic current induced in the system $J_{\mu}$ can be calculated
by differentiating the effective action, eq~(3.1), with respect to the
electromagnetic vector potential. In momentum space the induced current is
$$
J_{\mu}(\omega, {\vec Q}) = {1\over 2} {\tilde A}_{\nu}(-\omega, -{\vec Q})
[ K_{\mu\nu}(-\omega, -{\vec Q}) + K_{\nu\mu} (\omega, {\vec Q}) ] \eqno(4.3)
$$
In particular, the charge and the current density induced by the external
perturbation, eq ~(4.1), are
$$
J_{0}(\omega, {\vec Q}) =  {\tilde A}_{0}(-\omega, -{\vec Q})
       K_{00}(\omega, {\vec Q})  \eqno(4.4)
$$
$$
J_{j}(\omega, {\vec Q}) = {1\over 2} {\tilde A}_{0}(-\omega, -{\vec Q})
[ K_{j0}(-\omega, -{\vec Q}) + K_{0j} (\omega, {\vec Q}) ] \eqno(4.5)
$$
or in real space
$$
J_{0}({\vec x},t) =  2\pi {\tilde q}
                     \int {d^{2}Q\over (2\pi)^2}\;  |{\vec Q}|
                       {K_0(0,{\vec Q})} \; e^{i {\vec Q} { \vec x} }
\eqno(4.6)
$$
$$
J_{j}({\vec x}, t) =  2\pi {\tilde q}\; {\epsilon _{jk}} {\partial _{k}}
                           \int {d^{2}Q\over (2\pi)^2}
                           {K_1(0,{\vec Q}) \over |{\vec Q}|}
                           \; e^{i {\vec Q} { \vec x} }   \eqno(4.7)
$$
Using the expression (3.3) for $K_{0}$, the leading order term of the induced
charge becomes
$$
J_{0}( {\vec x},t) =-{{\bar \rho}{\tilde q}\over {M{\omega _c}^2}}\;
        { 1 \over |{\vec x}|^3} = -{\sigma _{xy}}\; {{\tilde q}\over
        {\omega _c}}\;{ 1 \over |{\vec x}|^3}      \eqno(4.8)
$$

Since the external perturbation generates an electric field in the radial
direction, the induced current given by eq ~(4.7) has only components in the
azimutal direction ($\hat \varphi$). The leading order term is
$$
 J_{\varphi}( {\vec x},t) =  {\sigma _{xy}}\; {\tilde q} \;
        { 1 \over |{\vec x}|^2}      \eqno(4.9)
$$
These results coincide with those obtained by Sondhi and Kivelson
\refto{sondhi}, who calculated the current induced by the presence of a
quasiparticle, within the framework of the Chern-Simons Landau-Ginzburg
theory for the FQHE.

The expressions obtained above are formally exact in the limit of infinite
magnetic field. Their corrections can be calculated by
taking into account higher order terms in the gradient expansions of the
functions $K_0$ and $K_1$, and in the semiclassical expansion.
These results hold if the pair potential ${\tilde V}({\vec Q})$ is
such that ${\vec Q}^2 {\tilde V}({\vec Q})$ vanishes when
${\vec Q}\rightarrow 0$. In both cases, for a short range potential or for the
Coulomb potential, the corrections to eq ~(4.8) and (4.9) will go as
${1\over |{\vec x}|^5}$ and ${1\over |{\vec x}|^4}$ respectively. In principle,
the corresponding coefficients might be renormalized by non-gaussian
fluctuations of the statistical gauge field.

We now calculate the linear response of the system in the presence of a static
magnetic flux located at the origin. We will consider here an
infinitesimally thin
flux tube with intensity $\Phi _{0}$, such that the system remains in
its ground state even in the
presence of the flux. If the flux through the solenoid gets to
be big enough, the system will be able to
lower its energy by creating quasiparticles or quasiholes (\ie\ moving
into an excited state),
which eventually might screen the flux. Hence, we expect that the
response
to an external infinitesimally thin solenoid of flux $\Phi _{0}$ should
be a periodic function of $\Phi _{0}$ with period equal to one flux
quantum. However, this problem cannot
be studied within the mean field solution that we have chosen, because
there is no way to go perturbatively from the uniform or liquid-like ground
state solution, to a solution which represents an excited state with one or
more quasiparticles present. To recover the expected periodic
behavior of the induced current as a function of the external flux,
we would have to study not only the uniform solution of the saddle point
equations, but also the vortex-like solutions, and sum over all of the
saddle points to obtain the full, periodic, response to an external
arbitrary flux. In this work we will only consider solenoids with flux
$\Phi_0$ much smaller than the flux quantum and, thus, we will only
consider the reponse of the uniform state.

The electromagnetic vector potential is in this case
$$
\eqalign{{\tilde A}_{0}({\vec x},t) &= 0 \cr
{\tilde A}_{j}({\vec x},t) & = {\Phi _{0} \over 2\pi |{\vec x}|}
             \;  {-\epsilon _{jk} x_k \over |{\vec x}|} \cr}    \eqno(4.10)
$$
or in momentum space
$$
{\tilde A}_{j}(\omega, {\vec Q}) = i \; 2\pi \delta (\omega ) \;  \Phi _{0}
               {\epsilon _{jk} Q_k \over |{\vec Q}|^2}    \eqno(4.11)
$$
According with eq~(4.3), the charge and current density induced by this
perturbation are, respectively
$$
J_{0}(\omega, {\vec Q}) = {1\over 2} {\tilde A}_{i}(-\omega, -{\vec Q})
[ K_{0i}(-\omega, -{\vec Q}) + K_{i0} (\omega, {\vec Q}) ] \eqno(4.12)
$$
$$
J_{j}(\omega, {\vec Q}) = {1\over 2} {\tilde A}_{i}(-\omega, -{\vec Q})
[ K_{ji}(-\omega, -{\vec Q}) + K_{ij} (\omega, {\vec Q}) ] \eqno(4.13)
$$
Substituting in these equations the explicit form of the external probe
(eq ~(4.11)), and tansforming back to real space, the induced charge and
current are given by
$$
J_{0}({\vec x},t) =  - \Phi _{0}\; \int {d^{2}Q\over (2\pi)^2}\;
                    {K_1 (0,{\vec Q})} \; e^{i {\vec Q}  {\vec x} } \eqno(4.14)
$$
$$
J_{j}({\vec x}, t) =  - \Phi _{0} \; {\epsilon _{jk}} {\partial _{k}}
                           \int {d^{2}Q\over (2\pi)^2}
                     K_2 (0,{\vec Q}) \; e^{i {\vec Q} { \vec x} }
\eqno(4.15)
$$
Keeping only the leading order terms in ${K_1 (0,{\vec Q})}$, eq~(3.4),
the induced charge becomes
$$
J_{0}(\vec x, t) = - K_{1}(0,0) \; \epsilon_{ik} \;
              \partial _{i} A_{k}(\vec x)   \eqno(4.16)
$$
where $K_{1}(0,0)$ is evaluated at zero frequency and momentum. Using that
$K_{1}(0,0) = {\nu \over 2\pi} $, we get
$$
J_{0}(\vec x, t) = - {\nu \over 2\pi}{\cal B}(\vec x)  \eqno(4.17)
$$
where ${\cal B}= \Phi _{0} {\delta ^{2} (\vec x) }$ is the the magnetic field
associated to the external flux, eq~(4.10).
It is important to remark that eq~ (4.17) is strictly valid in the limit
in which the external uniform magnetic field goes to infinity. Otherwise, we
find that the induced charge has gaussian factors that in the limit of infinite
magnetic field become delta fucntions which combine to reproduce exactly the
magnetic field produced by the external perturbation (eq~(4.10)).

The total charge $\tilde Q$ induced by the external
perturbation is
obtained by integrating eq~(4.17) over the area of the system. The result is
$$
{\tilde Q} = -\nu {\Phi _{0} \over 2\pi}      \eqno(4.18)
$$
Since the induced charge $\tilde Q$ has been determined from
linear response theory, it may seem that eq~(4.18) should only hold if
the flux $\Phi_0$ is small relative to the flux quanta. Eq~(4.18) is,
however, exact. This follows from the fact that the leading
behavior at small momentum of the response functions saturates the sum
rules and, in consequence, coefficients such as $K_1(0,0)$ are given
exactly by the linear response result. For instance,
if $\Phi _{0} = 2\pi$, the induced charge is just the filling fraction
of the
system. In particular, for $\nu = {1\over m}$ the induced charge is
$-{1\over m}$. This result agrees with Laughlin's {\it gedanken
experiment} argument for the construction of the quasihole.

Finally we consider the current induced by the solenoid.
If the interparticle pair potential is short range, beeing its range
much smaller than the magnetic length ($l$),  the induced current
density, eq~(4.15), is
$$
J_{\varphi}(\vec x, t) = ({\nu \over 2\pi})^2 \; \Phi _{0} \;
                           B_{\rm eff}\;  {\bar \omega }_{c}\;
                    |{\vec x}| \; e^{-{ B_{\rm eff} |\vec x |^{2} \over 2}}
                   \eqno(4.19)
$$
In particular, in the limit of $ B \rightarrow \infty$, the above expresion
becomes
$$
J_{\varphi}(\vec x, t) = -{\nu ^2 \over 2\pi M} \; \Phi _{0} \;
               {\partial \over \partial |{\vec x}|} {\delta ^2 (|\vec x |) }
                   \eqno(4.20)
$$
which is ultralocal.

If the pair potential has the Coulomb form, the leading order term in
the azimutal component of the induced current density is given by
$$
J_{\varphi}(\vec x, t) = ({\nu \over 2\pi})^2 \Phi_0
                     {q^2\over |{\vec x}|^2}       \eqno(4.21)
$$
which has a long range, power-law tail.

We close this section with a remark. The sum rule arguments
tell us that the asymptotic long distance behaviors for the charge and
current density profiles that we derived in this section, are exact.
However, at shorter distances they are expected to pick up corrections.
For instance, except for the case of the Coulomb
potential, our results show no dependence on the strength of the short
range potentials. This is so since we are considering {\it ultralocal}
pair interactions, \ie\ with a range comparable or smaller than the
cyclotron radius $l$. If, for example, we consider a
Yukawa-like interaction with a range $a$ much larger than $l$, but
still finite, we should expect a somewhat different behavior at
distances $R \approx a$. In fact, for $R \gg a \gg l$ we find
that the profiles decay exponentially fast with range $a$ ( \ie\ not
gaussians). For the regime $a\gg R \gg l$ we find, for the current
density profile, a $1/R$ power law decay  crossing over to
gaussian behavior at the scale of $l$. In other words, we only expect
changes either at the cyclotron scale or at any new length scales
introduced by the interaction.

\head{V. Application to the Interacting Anyon Gas}

In this section we consider a system of interacting anyons in the
absence of an external magnetic field.
This problem was previously discussed by many authors
\refto{fhl,chen,banks,e1,wiegman}. All of these works deal with anyons which,
apart from a hard core, are not interacting.

Here we find the spectrum of collective modes for different types of pair
interactions and we rederived some of
the previously known results on the electromagnetic response
functions within the framework of our theory.

For this problem, we can also expand the path integral $\cal Z$ around
stationary
configurations of the effective action. There are many possible solutions
 for the classical equations of motion, but we only study that one with uniform
particle density. At mean field level the anyons see a total flux $B_{\rm
eff}$,
which coincides with the average Chern-Simons flux,\ie
$B_{\rm eff} = - {{\bar \rho}\over \theta}$. In order for this theory to have a
gap in the single particle spectrum, an integer number $p$ of effective Landau
levels defined by $B_{\rm eff}$ must be completely filled. This requirement
implies a relation between $p$ and the statistics parameter given by ${p\over
2\pi} = - \theta$. Provided that this identity holds, the coefficient of
the
Chern-Simons term vanishes and the system has a gapless collective mode.

 The next step is to take into account the gaussian fluctuations. We can use
the results derived in section III for the electromagnetic response functions,
but with ${B_{\rm eff}}  = - {{\bar \rho}\over \theta}$
and ${p\over 2\pi} = - \theta$.
For the system of anyons we have also studied the spectrum of collective modes
only for the cases $p=1$ and $p=2$. The general case can be analized by using
the same methods.

\noindent{Case $p=1$}

Here $\theta = -{1\over 2\pi}$. Therefore the statistical angle is
$\delta = \pi$ and we are dealing with a system of interacting bosons.

We find that there is a family of collective modes whose zero-momentum
gap is $k {\bar \omega}_{c}$, where $k$ is an integer number different from
$1$, and whose dispersion curve $\omega_{k} ({\vec Q})$ is
$$\omega_{k} ({\vec Q})= \Big [ (k {\bar \omega}_{c})^2 -
                         ({{\vec Q}^2\over 2 {B_{\rm eff}}})^{k-1}\;
                         {\bar \omega}_{c}^2 \;
            {2 \over {(k-2)!}} \Big ] ^{1\over2} \eqno(5.1)
$$
The residue in $K_{00}$ corresponding to this pole is
$$
Res(K_{00},\omega _{k}({\vec Q})) = {\vec Q}^2 \;
                                   {{\bar \omega}_{c} \over 2\pi}\;
                                ({{\vec Q}^2\over 2 {B_{\rm eff}}})^{k-1}
                               {2 \over {k^{2} (k-2)!}}
                                  \eqno(5.2)
$$
In particular, if $k=2$, the above dispersion relation becomes
$$\omega_{2} ({\vec Q})= \Big [ ({4\pi {\bar \rho} \over M})^2 -
                         {v _0}^2 \; {\vec Q}^2 \Big ] ^{1\over 2}
\eqno(5.3)
$$
To obtain this expression we have used that ${{\bar \omega}_{c} \over 2\pi}=
{{\bar \rho} \over Mp}$. This mode appears to correspond to the
``density mode" of the Bose gas. Notice that all of the modes with $k
\geq 2$ are expected to become damped due to the non-quadratic interaction
terms which
induce decays into the gapless $k=1$ modes. The modes with larger
values of $k$, presumably, should get damped more quickly than the mode
at $k=2$.

The mode with $k=1$ is ``pulled down" and it becomes gapless (at ${\vec
Q}=0$). Its dispersion curve is
$$
\omega  ({\vec Q}) = {v _0} \; |{\vec Q}|      \eqno(5.4)
$$
where
$$
{v_0}^2 = { 2\pi {\bar \rho} \over {M^2}}  \eqno(5.5)
$$
Eq~(5.4) is valid provided that the pair potential is short range, and
that ${\tilde V}(0) = 0$.
If the pair potential is ${\tilde V}(\vec Q) = {2\pi q^2 \over
|{\vec Q}|}$, the gapless mode has the form
$$
\omega ^{2} ({\vec Q})={v _0}^2 \;Mq^2\; |{\vec Q}|         \eqno(5.6)
$$
In both cases the residue is
$$
Res(K_{00},\omega ({\vec Q})) = -{\vec Q}^2  \; {{\bar \rho} \over M}
                                  \eqno(5.7)
$$
These results agree with the general discussion by Fetter \refto{fetter} of
plamons in two-dimensional compressible fluids. This behavior is a consequence
of the compressibility and \hfill \break
two-dimensionality and it is independent of the
statistics of the particles.

\noindent{Case $p=2$}

Here $\theta = -{2\over 2\pi}$. This is the case of semions.
Again we find that there is a family of collective modes whose zero-momentum
gap is $k {\bar \omega}_{c}$, where $k$ is an integer number different from
$1$, and whose dispersion curve $\omega_{k} ({\vec Q})$ is
$$\omega_{k} ({\vec Q})= \Big [ (k {\bar \omega}_{c})^2 -
                         ({{\vec Q}^2\over 2 {B_{\rm eff}}})^{k-1}\;
                         {\bar \omega}_{c}^2 \;
            {(k+2) \over {(k-2)!}} \Big ] ^{1\over2} \eqno(5.8)
$$
The residue in $K_{00}$ corresponding to this pole is
$$
Res(K_{00},\omega _{k}({\vec Q})) = {\vec Q}^2 \;
                                   {{\bar \omega}_{c} \over 2\pi}\;
                                ({{\vec Q}^2\over 2 {B_{\rm eff}}})^{k-1}
                               {(k+2) \over {k^{2} (k-2)!}}
                                  \eqno(5.9)
$$
In particular, if $k=2$, the above dispersion curve coincides with
eq~(5.3).
Again, the mode at $k=2$ can be regarded as a ``density mode". All
of these modes will also become damped by interaction effects.

The mode with $k=1$ is ``pulled down" and it becomes gapless (at ${\vec
Q}=0$).
Its dispersion curves and residue are the same as in the case $p=1$,
eq~(5.4), ~(5.6) and ~(5.7) respectively.

We have further checked that for any other value of $p$, the gapless mode and
the mode with zero momentum gap at $2{\bar \omega _{c}}$ have the same form
as those for $p=1$, eq~(5.3), ~(5.4), and ~(5.6).

We have seen then that the electromagnetic response functions that we find at
the semiclassical level have a gapless collective excitation. We can see
also that there is a restoration of parity in the uniform limit, and that the
system exhibit Meissner effect. These two last results are already well known,
but we reproduce them here for completeness.

For short ranged pair interactions, the density correlation function is,
to leading order in ${\vec Q}^2$ $$
K_{00}(\omega, {\vec Q}) = - {{\bar \rho}\over M} \;
  {{\vec Q}^2 \over {\omega }^2 - {v_0}^2 {\vec Q}^2} \eqno(5.10)
$$
Therefore, $K_{00}$ has a massless pole which corresponds to a gapless
collective
mode whose velocity is given by ${v_0}$. In this sense we can say that the
Gaussian fluctuations {\it restore} the  compressibility of the system.
This coincides with the results of  references \cite{chen} and \cite{wiegman}.

The coefficient of the Chern-Simons term, eq~(3.4), has the following
properties
$$
{\lim _{\omega \rightarrow 0}}{ K_1}(\omega,{\vec Q}) =-\frac{p}{8 \pi}
\eqno(5.11)
$$
if the limit $\omega \to 0$ is taken first. However, if $Q^2 \to 0$
first, we find
$$
{\lim _{{\vec Q} \rightarrow 0}}{ K_1}(\omega,{\vec Q}) =0  \eqno(5.12)
$$
Eq~(5.12) simply means that the Hall conductance of the anyon gas is
zero. Halperin \etal\ \refto{halperinetal} have argued that this result
is a consequence of
Galilean invariance. Eq~(5.11) can be thought as a static response of
the ground state to a periodic modulation of the charge density with
wave vector ${\vec Q}$ which induces a periodic arrangement of currents.
These currents give rise to an orbital magnetic moment. Eq~(5.11) is the
static (or equilibrium) orbital suceptibility. It is hard to
believe that for the case of bosons, which do not have any violation of
time reversal, there should be any orbital currents. These results
coincide with those of FHL\refto{fhl} and Chen
\etal \refto{chen}, at the Hartree level.
Dai \etal
\refto{fhl} have shown recently that non-gaussian fluctuations (beyond
the RPA) yield a limiting value of zero for $K_1(\omega, {\vec Q})$ (as
$\omega \to 0$) for bosons but not for semions.

Finally, we will show that the system exhibit Meissner effect.
In the limit of long distances, and for short range interactions, the
coefficient $K_2$, eq~ (3.5), can be written as
$$
K_2 (\omega , {\vec Q}) = - {2\pi {\bar \rho}\over {M^2}} \;
                 K_{0} (\omega , {\vec Q})  \eqno(5.13)
$$
where according to eq~(3.2) and ~(5.10) $K_0$ is
$$
K_{0}(\omega,{\vec Q}) = - {{\bar \rho}\over M} \;
  {1 \over {\omega }^2 - {v_0}^2 {\vec Q}^2} \eqno(5.14)
$$
The electromagnetic current induced in the system because of the presence of a
magnetic field is $J_k = K_{kj} A_j$. Using eq~(3.2) and ~(5.12) we get
$$
J_k = ({\omega}^2  K_{0}(\omega, {\vec Q})
   +{\vec Q}^2 K_{2}(\omega, {\vec Q})) A_k
   - {Q_k}{Q_j}K_{2}(\omega, {\vec Q}) A_j    \eqno(5.15)
$$
Therefore, the curl of the current is, in momentum space
$$
\epsilon_{lk} \;  Q_l J_k = -{{\bar\rho }\over M}
                \epsilon_{lk} \;  Q_l A_k  \eqno(5.16)
$$
This is precisely the London equation, where the London penetration depth is
${1\over \lambda^2} = {4\pi {\bar \rho} \over M}$.

\head{VI: Conclusions}

In reference \cite{lopez1} we developed a Chern-Simons theory for the FQHE
based on a second-quantized fermion path-integral approach.
In this paper we have calculated the electromagnetic response functions of the
Fractional Quantum Hall system within the framework of that theory.
We made a semiclassical expansion and we worked around the average field
approximation. As we have already mentioned above, the mean field
solution
violates explicitly Galilean invariance. At this level of the approximation,
the center of mass of the system executes a cyclotron-like motion at the
effective cyclotron frequency. In this sense the gaussian (or semiclassical)
fluctuations are essential to restore the original symmetries of the problem.
We saw that order-by-order in the semiclassical expansion the
response functions obey the correct symmetry properties required by
Galilean and Gauge invariance, and by the incompressibility of the fluid.
We showed that, already at the semiclassical or gaussian level, the
low-momentum limit of the density correlation function saturates the
{\it f}-sum rule, and in that sense this result is exact, \ie\ it can not be
renormalized by higher order corrections. We calculated the Hall conductance
out of the density-current correlation functions, and we found that it has
the correct value at this order of the approximation.
We obtained the spectrum of collective excitations in the low-momentum limit
for short-range and for Coulomb interparticle pair potential.
We found that there is a family of collective modes whose zero-momentum
gaps are integer multiples of the effective cyclotron frequency. In particular,
there are two modes merging at the cyclotron frequency at zero momentum, but
with different intensities, \ie\ different weights in their residues in the
density correlation function. We argued that all of these modes, except the one
with least energy, will be damped once the higher order terms in the
semiclassical expansion are taken into account.
We also calculated the linear response of the system to external charges and
fluxes, and found expressions for the asymptotic form of the charge and
current density profiles. We found that the responses to an external
charge always show profiles with universal power law decays. In
contrast, the responses
to external infinitesimally thin solenoids exhibit a variety of
behaviors which depend on the nature of the interactions.

Finally, we reconsidered the anyon superfluid within our scheme and derived the
spectrum of collective excitations for interacting hard-core bosons and
semions. In addition to the gapless phase mode, we found a set of gapped
collective modes.

There are still many questions left open. The theory presented here
provides a
good description of the uniform FQHE ground state, of its collective
excitations, and of its linear electromagnetic response. We have made a
number of predictions about the existance of a family of collective
modes. The
observability of them depends not only on their intensities, but also on their
life-times. We have not addressed this problem here.
Another open point of interest is the study of this theory starting from
another
solution of the saddle-point equations, as the Wigner crystals and non-uniform
states with vortex-like configurations.

\head{Acknowledgements}

We thank Patrick Lee for an early discussion on his work on the
even denominator states and to Bert Halperin, Patrick Lee and Nick Read
for making their paper available to us prior to publication. We also
thank Mike Stone for many stimulating conversations on the subject.
We are very grateful to Shivaji Sondhi for a very useful comment on the
response to an external flux. This work was supported in part by NSF
Grant No.DMR91-22385 and by the
Science and Technology Center for Superconductivity at the University of
Illinois at Urbana-Champaign through the grant DMR88-09854/21. We are
grateful to Prof.~M.~Klein for his encouragement and support.

\endpage

\references

\refis{wiegman} P.B. Wiegman, \prl 65, 2070, 1990.

\refis{banks} T. Banks and J.D. Lykken, \journal Nucl. Phys. B, 36,
500, 1990.

\refis{chen} Y.H. Chen, B.I.Halperin, F. Wilczek, and E. Witten,
\journal Int. J. Mod. Phys. B, 3, 1001, 1989.

\refis{lopez1} A. Lopez and E. Fradkin , \prb 44 , 5246, 1991.

\refis{jain1} J.Jain, \prl 63, 199, 1989, \prb 40, 8079, 1989, and
\journal Advances in Phys., 41, 105, 1992.

\refis{wilczek1} F.Wilczek, \prl 48, 1144, 1982.

\refis{book} E.Fradkin, {\it ``Field Theories of Condensed Matter
Systems"}, Addison-Wesley, Redwood City (1991).

\refis{sondhi} S.L. Sondhi and S.A. Kivelson, UCLA preprint, 1991.

\refis{e1} E.Fradkin, \prb 42, 570, 1990.

\refis{fhl} C.B. Hanna, R.B. Laughlin and A.L. Fetter, \prb 40, 8745, 1989;
and \prb 43, 309, 1991; Q. Dai, J.L. Levy, A.L. Fetter, C.B. Hanna and
R.B. Laughlin, Stanford prepint, (1992).

\refis{klein} M.V. Klein in {\it ``Light Scattering in Solids I"}, M.
Cardona Editor, Springer-Verlag, New York (1983).

\refis{kohn} W. Kohn, \journal Physical Review, 123, 1242, 1961.

\refis{dreamteam} B.~I.~Halperin, P.~A.~Lee and N.~Read, MIT preprint,
July 1992.

\refis{lopez2} A. Lopez and E. Fradkin , \prl 69, 2126, 1992.

\refis{ramond} To be more precise, whenever we have had to
fix the gauge we have adopted Feynman's method of averaging
over gauges, with $\alpha$ being the width of the
distributions. This is a standard procedure which is reviewed in many
textbooks. See, for instance, P.~Ramond,
``Field Theory; A Modern Primer", Addison-Wesley, Redwood City (1989).

\refis{fetter} A.~L.~Fetter, \prb 10, 3739, 1974; \journal Annals of
Physics (NY), 81, 367, 1973.

\refis{zhang} S.C.Zhang, T.Hansson, and S. Kivelson, \prl 62, 82, 1989.

\refis{zhang1} S.C.Zhang, \journal Int. Jour. Mod. Phys. B, 6, 25, 1992.

\refis{kallin} C.~Kallin and B.~I.~Halperin, \prb 30, 5655, 1984.

\refis{girvin2} S.Girvin, A.MacDonald and P.Platzman, \prl 54, 581,
1985; \prb 33, 2481, 1986.

\refis{puzzle} The splitting at $\nu=1/3$ appears as follows. Eq~(3.12)
and eq~(3.14) show that, for general $m$, the dispersions of the modes
$\omega_{\pm} ({\vec Q})$ go like ${\vec Q}^2$ and ${\vec Q}^{2(m-2)}$
respectively. For $\nu=1/3$ (\ie\ $m=3$) they both disperse like ${\vec
Q}^2$. From the point of view of the expansion in powers of ${\vec
Q}^2$, the cyclotron mode, in general, has contributions only from the
first
two leading poles of $\Pi_0 $, $\Pi_1$ and $\Pi_2$. But, for $m=3$, the
third leading pole gives a contribution with the same order in ${\vec
Q}^2$. This additional contribution is responsible for this unusual
behavior.

\refis{laughlin1} R.B.Laughlin, \prl 50, 1395, 1983, and in
{\it ``The Quantum Hall Effect"},R.Prange and S.Girvin Editors,
Springer-Verlag, New York (1989).

\refis{hald} F.D.M.Haldane, \prl 51, 605, 1982, and in
{\it ``The Quantum Hall Effect"},
R.Prange and S.Girvin Editors, Springer-Verlag, New York (1989);

\refis{halp} B.I.Halperin, \journal Helv. Phys. Acta, 56, 75, 1983,
and \prl 52, 1583, 1984.

\refis{halperinetal} B.~I.~Halperin, J.~March-Russell and F.~Wilczek,
\prb 40, 8726, 1989.

\refis{pinczuk} A.~Pinczuk, J.~P.~Valladares, D.~Heiman, A.~C.~Gossard,
J.~H.~English, C.~W.~Tu, L.~Pfeiffer and K.~West , \prl 61, 2701, 1988;
see also B.~B.~Goldberg, D.~Heiman, A.~Pinczuk, L.~Pfeiffer and K.~West,
\journal Surface Science, 263, 9, 1992.

\endreferences

\endit

%%%%%%%%%%%%%%%%%%%%%%%%%%%%%%%%%%%%%%%

\refis{tsui} D.C. Tsui, H.L. Stormer, and A.C. Gossard, \prl 48, 1559, 1982.

\refis{daniel} D.~Boyanovsky, University of Pittsburgh preprint (1992).

\refis{galilean} B.~I.~Halperin, J.~March-Russell and F.~Wilczek, \prb
40, 8726, 1992.

\refis{iengo} R.Iengo and K.~Lechner, \np {\bf B}346, 551, 1990;
R.Iengo and K.~Lechner, to be published in {\it Physics Reports}.

\refis{ef} E. Fradkin, University of Illinois preprint (1992).

\refis{chang} A.M. Chang in {\it ``The Quantum Hall Effect"},
R.Prange and S.Girvin Editors, Springer-Verlag, New York (1989);
T.Sajoto, Y.W. Suen, L.W. Engel, M.B.Santos, and M. Shayegan,
\prb 41, 8449, 1990 and the references cited therein.

\refis{willet} R.L. Willet, J.P. Eisenstein, H.L. Stormer, D.C. Tsui,
A.C. Gossard, and J.H. English  \prl 59, 1776, 1987.

\refis{girvin1} S.Girvin and A.MacDonald, \prl 58, 1252, 1987.

\refis{read} N.Read,  \prl 62, 86, 1989.

\refis{kt} S.Trugman and S.Kivelson, \prb 31, 5280, 1985.
constant $\theta$.